\documentclass[aps,prl,twocolumn]{revtex4}

\usepackage{graphicx}

\usepackage{times}
\usepackage{amsmath}
\newcommand{\amu}{a_\mu}
\newcommand{\tb}{\tan\beta}
\newcommand{\tbsq}{\tan^2\beta}

\newcommand{\amuSUSY}{a_\mu^{\rm SUSY}}

\newcommand{\amuSUOL}{a_\mu^{\rm SUSY,1L}}

\newcommand{\amuSUTLferm}{a_\mu^{\rm SUSY,ferm,2L}}
\newcommand{\amuSUTLbos}{a_\mu^{\rm SUSY,bos,2L}}

\newcommand{\SLASH}[2]{\makebox[#2ex][l]{$#1$}/}

\newcommand{\pslash}{\SLASH{p}{.2}}

\newcommand{\ov}{\overline}
\newcommand{\gev}{\mbox{GeV}}
\newcommand{\Tan}{\mbox{Tan}\,}

\begin{document}
\title{\boldmath $\Tan\beta$-enhanced supersymmetric corrections to the
  anomalous magnetic moment of the muon\unboldmath} \author{Schedar
  Marchetti$^1$} \author{Susanne Mertens$^1$} \author{Ulrich Nierste$^1$}
\author{Dominik St\"ockinger$^2$}

\affiliation{~\\
       $^1$
       Institut f\"ur Theoretische Teilchenphysik (TTP)\\
        Karlsruhe Institute of Technology --- Universit\"at Karlsruhe\\
        76128 Karlsruhe, Germany\\ 
       $^2$
Institut f\"ur        Kern- und Teilchenphysik,\\
TU Dresden, D-01062 Dresden, Germany
}

\begin{abstract}
  We report on a two-loop supersymmetric contribution to the magnetic moment
  $(g-2)_\mu$ of the muon {which is enhanced by two powers of
    $\tan\beta$.  This contribution arises from a shift} in the relation
  between the muon mass and Yukawa coupling and can {increase} the
  supersymmetric contribution to $(g-2)_\mu$ {sizably}. As a result,
  if 
  the currently observed $3\sigma$ deviation between the experimental and 
  SM theory value of $(g-2)_\mu$ {is analyzed within the 
  Minimal Supersymmetric Standard Model (MSSM), the derived 
  constraints on the parameter space 
  are modified significantly:
If  $(g-2)_\mu$ is used to determine  $\tan \beta$ as a
function of the other MSSM parameters, our corrections decrease 
$\tan \beta$ by roughly 10\% for $\tan \beta =50$.} 
\end{abstract}
\maketitle
The anomalous magnetic moment $\amu=(g-2)_\mu/2$ of the muon is one of
the most precisely measured and calculated quantities in particle
physics --- and recently it has developed into one of the observables
with the {most significant} deviations between the experimental
value and the corresponding Standard Model (SM) theory prediction. The
review \cite{MillerRR} obtains%
\begin{align} \amu^{\rm exp}-\amu^{\rm SM} &= 29.5(8.8)\times10^{-10},
\label{deviation}
\end{align}
a $3.4\sigma$ deviation between $\amu^{\rm exp}$ and $\amu^{\rm SM}$,
the experimental \cite{BNL6} and SM theory value, respectively.

Eq.\ (\ref{deviation}) represents dramatic progress. It has been 
made possible by better determinations of the hadronic $e^+e^-$ cross 
section by SND, CMD-II, KLOE and
BaBar \cite{experiments}.
These are crucial ingredients for all recent evaluations
of the hadronic vacuum polarization contribution to $\amu^{\rm SM}$
\cite{Jegerlehner07,Davier07,HMNT06}.
{In Ref.\ \cite{Benayoun}, further progress on the
``$\tau$-puzzle'' has been achieved, confirming the $e^+e^-$-based result
(\ref{deviation}).} Now all evaluations of the SM theory prediction have a
smaller error than ever before and agree very well. 
{With this} 
progress 
the case for physics beyond the SM in $\amu$ has become stronger.
{Generically,} contributions from {new} physics with characteristic
mass scale $M_{\rm BSM}$ are suppressed as $(M_W/M_{\rm BSM})^2$ compared to
the SM electroweak contribution {of $\amu^{\rm
    weak}=15.4(0.2)\times10^{-10}$}, which is only half as large as the
observed deviation. {Thus some parametric enhancement of the new
  contribution is required.} 

Supersymmetry (SUSY), implemented in the Minimal
Supersymmetric Standard Model (MSSM), {can naturally explain the}
observed deviation {for two reasons:} 
First, the masses of smuons and charginos, the most relevant SUSY
particles, can be as small as $M_{\rm 
  SUSY}\sim{\cal O}($100~GeV$)$ without contradicting current
experimental data, {allowing a rather mild} suppression 
factor $(M_W/M_{\rm   SUSY})^2$. 
Second, the SUSY contributions to $\amu$ are enhanced by the parameter
\begin{align}
\tan\beta&=\frac{v_2}{v_1},
\end{align}
the ratio of the vacuum expectation values (vevs) of the two Higgs doublets
$H_{1,2}$ in the MSSM, {which governs the size of the down-type
  Yukawa couplings. We normalize the vevs as 
  $v\equiv\sqrt{v_1^2+v_2^2}=174\; \gev$.
  Since $a_{\mu}$ involves a chirality-flip, it
  is proportional to the muon Yukawa coupling $y_{\mu}$.}  
Large values $\tan\beta\sim50$ lead to similar top and bottom Yukawa
couplings and are therefore preferred {in scenarios with} Yukawa
unification. {Remarkably, naive multi-Higgs doublet models fail to 
explain the deviation in Eq.~(\ref{deviation}), because the corresponding 
loop diagrams involve at least three powers of the small coupling 
$y_{\mu}$.} 

The SUSY contributions to $\amu$ are approximately given by 
\begin{align}
\amu^{\rm SUSY}\approx
13 \times10^{-10}\left(\frac{100\,\rm GeV}{M_{\rm SUSY}}\right) ^2
\ \tan\beta\ \mbox{sign}(\mu),
\label{approx}
\end{align}
if the SUSY parameters for the smuon, gaugino and Higgsino masses have
a common scale $M_{\rm SUSY}$, see \cite{review} and
references therein. The sign of the
contributions is given by the sign
of the Higgsino mass parameter $\mu$ {(choosing the gaugino mass parameters
$M_1$, $M_2$ positive)}. {We restrict our analysis to the case of real
  $\mu$, $M_1$ and $M_2$, 
  because sizable CP-violating phases of these parameters are in
  conflict with the bounds on electric dipole moments, if $M_{\rm SUSY}$
  is in the range needed to accommodate $\amu^{\rm exp}$.} 
The SUSY contributions explain the entire deviation of
$29.5\times10^{-10}$ if 
{$\tan\beta$ is given by approximately $ 2.3  ({M_{\rm SUSY}}/{100\,\rm GeV}) ^2$.}

{Clearly, $\amu$ plays an eminent role in studies of the MSSM parameter
  space, see e.g.\ \cite{scanrefs}. 
In Ref.~\cite{Ellis} a preference of a constrained version of the
  MSSM over the SM is found from a global fit to collider and electroweak
  precision data. {This conclusion is primarily driven by
    $\amu$. In Refs.\ \cite{Hertzog:2007hz,PlehnRauch} a 
    future $\tan\beta$-determination using LHC-data combined with
    $\amu$ is outlined. Owing to its importance, $\amu$
  therefore deserves a theoretical precision analysis including radiative
  corrections.}}

So far, all
SUSY one-loop contributions are known \cite{oneloop}. 
At the two-loop level two kinds of {relevant} SUSY contributions 
{have been identified:}
QED-logarithms  $\log(M_{\rm SUSY}/m_\mu)$ arising from SUSY
one-loop diagrams with additional photon exchange have been evaluated in
\cite{DG98} and amount to $-7$\% to $-9$\% of the one-loop
contributions. Two-loop diagrams involving closed loops of either
sfermions (stops, sbottoms, etc) or charginos/neutralinos have been
evaluated in \cite{HSW0304}. They amount to about $2\%$ of the one-loop
contributions if all SUSY masses are degenerate but can be much larger,
if e.g.\ smuon masses are very heavy but stops and/or 
charginos and Higgs bosons are light.

All these known SUSY contributions to $\amu$ share the feature of 
{Eq.~}(\ref{approx}): For large $\tan\beta$ they are linear in
$\tan\beta$,
\begin{align}
\label{tborder}
\amu^{\rm SUSY,\ known} \propto \alpha^l
\left(\frac{m_\mu}{M_{\rm SUSY}}\right)^2\tan\beta,
\end{align}
where $l=1,2$ denotes the loop order.
In this paper we identify and discuss a SUSY contribution $\amu^{\rm
SUSY,\ \Delta_\mu}$ which is quadratic in $\tan\beta$, i.e.\ of the order
\begin{align}
\label{tbsqorder}
\amu^{\rm SUSY,\ \Delta_\mu} \propto \alpha^2 
\left(\frac{m_\mu}{M_{\rm SUSY}}\right)^2\tbsq,
\end{align}
and can therefore be a significant correction in the large-$\tan\beta$
region.

The physical origin of these $\tbsq$-corrections is a shift in the muon Yukawa
coupling $y_\mu$ due to $\tan\beta$-enhanced one-loop effects. In the
computation of $\amu^{\rm SUSY}$ beyond the one-loop level this shift appears
in the muon mass renormalization constant $\delta m_\mu$, defined in the
on-shell scheme:
\begin{align}
m_\mu+\delta m_\mu &=
\frac{m_\mu}{1+\Delta_\mu}+\mbox{non-}\tan\beta\mbox{-enhanced terms},
\nonumber\\
\label{shift}
y_\mu &= \frac{m_\mu}{v \cos \beta (1+\Delta_\mu)} 
         \left( 1 + {\cal O} (\cot \beta) \right),
\end{align}
where $m_\mu$ is the physical, pole-mass of the muon and where the
{shift $\Delta_\mu\propto\alpha\tan\beta$ will be given below}. 
{This type of
  $\tan\beta$-enhanced corrections has been studied intensely in the
  down-quark sector \cite{tanbe,CGNWlong}. In the standard approach one
  employs the limit $M_{\rm SUSY}\gg v_2$ and derives an effective
  loop-induced coupling of $H_2$ to down-type fermions, which results in
  relations between masses and Yukawa couplings of the type in
  Eq.~(\ref{shift}) \cite{tanbe}.  For $\amu$, however, this procedure fails,
  because $\amu^{\rm SUSY}$ vanishes in the limit $M_{\rm SUSY}\gg v_2$, so
  that the important corrections associated with $\Delta_\mu$ were overlooked
  so far. In the case of $\amu$ one must resort to the method of
  Ref.~\cite{CGNWlong}, which explicitly identifies $\tan\beta$-enhanced loop
  diagrams and resums them to all orders in perturbation theory for the 
  case of interest $M_{\rm SUSY}\sim v_2$. Eq.~(\ref{shift}) contains
  the desired effect to all orders $\alpha^l \tan^l\beta$,
  $l=1,2,\ldots$. For the phenomenology of $\amu$ only the term with
  $l=1$, contributing to $\amu^{\rm SUSY}$ at the two-loop level, is 
  relevant.}

In the following we will show that the shift {in Eq.~}(\ref{shift})
is the only source of the {$\tan\beta$-enhanced radiative
  corrections of type $\alpha^l \tan^l\beta$ and that there are no
  enhancement factors with even more powers of $\tan\beta$.}  The proof
relies on an analysis of mass singularities similar to the analysis
presented in \cite{CGNWlong}: {The one-loop diagram proportional to
  $y_\mu$ gives one power of $m_\mu \tan\beta$. (The second factor of
  $m_\mu$ in Eqs.~(\ref{tborder}) and (\ref{tbsqorder}) stems from the
  definition of $\amu$.)  A genuine $l$-loop diagram (i.e.\ without
  counterterms) may involve $n$ powers of $\tan \beta$ stemming from the
  muon Yukawa coupling $y_\mu$ or any other Yukawa coupling $y_f\propto
  (m_f/M_W) \tan \beta$. It will result in a desired
  $\tan\beta$-enhanced correction if $n\geq l$ and the loop diagram
  diverges as $1/(m_f^{n-1})$ for $m_f\to 0$ to compensate for the
  factor of $m_f^n$ in $y_f^n$. 

Such mass singularities can be analyzed
  by passing from the MSSM to an effective field theory in which all
  heavy particles are integrated out and only particles with mass $m_f$
  or less are retained. Heavy loops are represented by point-like
  interactions in the effective theory and the infrared structure of any
  MSSM loop diagram and its counterpart in the effective theory are the
  same. A novel feature compared to the analysis of Yukawa interactions
  in Ref.~\cite{CGNWlong} is the appearance of one dimension-5 coupling,
  the magnetic interaction term $\ov\mu_L \sigma_{\nu\rho} \mu_R
  F^{\nu\rho}$. On dimensional grounds any loop corrections involving
  this term can only depend logarithmically on light fermion masses
  $m_f$. Potentially dangerous loops involve effective couplings of
  dimension 4 or less, since they might come with one or more inverse
  power of $m_f$. However, the only such couplings induced by heavy
  loops are those which are already present in low-energy QED and QCD
  and the effect of the heavy particles in the underlying theory can be
  completely absorbed into the renormalization of masses and couplings
  in the effective theory \cite{ac}. 

In conclusion the only effective
  diagrams with inverse powers of $m_f$ are the known QED diagrams
  proportional to $1/m_\mu$ and they are unaffected by our MSSM
  short-distance structure.  These findings only hold, if a decoupling
  scheme is adopted for the renormalization \cite{ac}; for our case it
  is important that $m_\mu$ is renormalized in the on-shell scheme. Next
  we inspect other diagrams involving counterterms: The counterterm for
  the Yukawa coupling $y_\mu$ is $\delta y_\mu= y_\mu \delta m_\mu/m_\mu
  \propto m_\mu \tan^2 \beta$ and gives rise to the enhanced corrections
  in Eq.~(\ref{shift}) \cite{CGNWlong}. The counterterms for gauge
  couplings and the muon and photon fields cannot be
  $\tan\beta$-enhanced, because unlike $\delta y_\mu/y_\mu$ they do not
  involve any factor of $1/m_f$.  Finally the renormalization of the
  Higgs\-ino mass parameter $\mu$, the soft SUSY breaking terms and the
  parameter $\tan\beta$ can be chosen at will, and our statement about
  the absence of $\tan\beta$-enhanced corrections beyond those in
  Eq.~(\ref{shift}) is valid} for all renormalization schemes in which
{these} renormalization constants are {\em not}
$\tan\beta$-enhanced.  This includes common schemes such as the
$\overline{\mbox{DR}}$-scheme for all SUSY parameters, or a mixed scheme
where $\tan\beta$ and the $A$-parameter are defined in the
$\overline{\mbox{DR}}$-scheme but the {smuon, chargino and
  neutralino masses are renormalized on-shell.}

The shift $\Delta_\mu$ is given by the $\tan\beta$-enhanced terms of the muon
self energy. In terms of the loop function
\begin{align}
 I(a,b,c)&=\frac{
  a^2b^2\log{\frac{a^2}{b^2}}
        +b^2c^2\log{\frac{b^2}{c^2}}
        +c^2a^2\log{\frac{c^2}{a^2}}}{(a^2-b^2)(b^2-c^2)(a^2-c^2)}
,
\end{align}
which satisfies  $I(a,a,a)=1/(2a^2)$, it can be written as
\begin{align}
\Delta_\mu =
&-{\mu \ \tan\beta\ }
\frac{g_2^2\ M_2}{16\pi^2}\ I(m_1,m_2,m_{{\tilde{\nu}_\mu}})\nonumber\\
&
-{\mu \ \tan\beta\ }
\frac{g_2^2\ M_2}{16\pi^2}\ \frac{1}{2} I(m_1,m_2,m_{\tilde{\mu}_L})
\nonumber\\
& -{\mu \ \tan\beta\ } \frac{g_1^2\ M_1}{16\pi^2}\
\Big[
I(\mu,M_1,m_{\tilde{\mu}_R})\nonumber\\
& -\frac{1}{2}
I(\mu,M_1,m_{\tilde{\mu}_L})
-I(M_1,m_{\tilde{\mu}_L},m_{\tilde{\mu}_R})\Big].
\label{DeltamuResult}
\end{align}
The appearing gaugino, Higgsino, and smuon mass
parameters {and the Standard Model parameters $g_{1,2}$, $s_W$ are
  defined as usual, see e.g.\ \cite{review}}, and we have defined
\begin{align}
m_{1,2}^2 &=\frac{1}{2}\Big[
(M_2^2 + \mu^2 + 2M_W^2)\nonumber\\ 
&\qquad\mp\sqrt{(M_2^2 + \mu^2 + 2M_W^2)^2-4M_2^2\mu^2}\Big], \nonumber \\
m^2_{\tilde{\nu}_\mu}& 
=m_{L,\tilde{\mu}}^2- \frac{M_Z^2}{2},\quad
m^2_{\tilde{\mu}_L} = m_{L,\tilde{\mu}}^2-M_Z^2 (s_W^2-\frac12),
\nonumber \\
m^2_{\tilde{\mu}_R}&=m_{R,\tilde{\mu}}^2+M_Z^2 s_W^2.
\end{align}
While the chargino contributions are exact in the large-$\tan\beta$
limit, the neutralino 
contributions in (\ref{DeltamuResult}) have been simplified using the
approximation $M_Z\ll \mu, M_1, M_2$.  {The} deviation of
$\Delta_\mu$ as given in (\ref{DeltamuResult}) from the exact result
{satisfies} $|\Delta_\mu-\Delta_\mu^{\rm exact}|<0.01$ over the entire 
parameter range (all supersymmetry masses {are varied} independently
between $100$ GeV and $2$ TeV, $\tan\beta\le100$) for which
$|\amu^{\rm SUSY}|<10^{-8}$.
  
Similar shifts exist for all down-type fermions, and in particular the
shift of the bottom-quark Yukawa coupling $\Delta_b$ has been analyzed
in detail in the literature \cite{tanbe,CGNWlong}, 
and the results can be readily applied to the muon case.

{The contribution $\amu^{\rm SUSY,\Delta_\mu}$ of the new $\tb$-enhanced
  contributions to {$\amu=-2m_\mu F_M(0)$ can be easily
    obtained by noting that the magnetic form factor $F_M(0)$} is
  proportional to $y_\mu$, apart from numerically irrelevant terms with three
  or more powers of $y_\mu$. Now $y_\mu$ enters $F_M$ in two ways: First it
  appears explicitly in the higgsino-muon couplings or in the Higgs-smuon
  coupling triggering the left-right mixing in the smuon mass matrix. Second
  it appears implicitly through $m_\mu \propto y_\mu \cos\beta$, which 
  arises from the application of the Dirac equation 
  $\pslash \mu_L=m_\mu \mu_R$. The second contribution is suppressed by a
  factor of $\cot\beta$ compared to the first. The $\tb$-enhanced
  corrections to the first  
  contribution are obtained by using the expression in Eq.~(\ref{shift}) 
  for $y_\mu$. Therefore}
\begin{align}
\amu^{\rm SUSY,1L}+\amu^{\rm SUSY,\Delta_\mu} &=
\amu^{\rm SUSY,1L}\left(\frac{1}{1+\Delta_\mu}\right).
\label{ResultamuDeltamu}
\end{align}
Note that {this formula is} only correct for the 
{enhanced terms of order $\alpha^l \tan^l\beta$},
but this is sufficient for our purposes.

Equation (\ref{ResultamuDeltamu}) is the main result of this paper. We are now
in the position to write down the most accurate prediction for $\amu^{\rm
  SUSY}$, replacing the result given in \cite{review} {by}
\footnote{{Note that there is no double-counting between the
    non-$\tan^2\beta$-enhanced terms implicitly contained in
    (\ref{ResultamuDeltamu}) and the terms in the second and third
    line of (\ref{amuSUSYknown}).}}
\footnote{In our numerical analysis we parametrize the one-loop result
  in terms of the muon decay constant $G_\mu$, i.e.\ we replace
  $\pi\alpha/s_W^2\to \sqrt2 G_\mu M_W^2$, in order to absorb further
  universal two-loop corrections. This gives rise to slight
  numerical differences compared to \cite{review}.}
\begin{align}
\amuSUSY &=
\amuSUOL\left(1-\frac{4\alpha}{\pi}\log\frac{M_{\rm
    SUSY}}{m_\mu}\right)\left(\frac{1}{1+\Delta_\mu}\right)\nonumber\\
&+
\amu^{(\chi\gamma H)}+\amu^{(\tilde{f}\gamma H)}
+\amu^{(\chi\{W,Z\} H)}+\amu^{(\tilde{f}\{W,Z\}H)}\nonumber\\
&
+\amuSUTLferm + \amuSUTLbos
+\ldots.
\label{amuSUSYknown}
\end{align}
The first line contains the one-loop result, corrected by large QED-logarithms
\cite{DG98} and by the {new $\tb$-enhanced terms} discussed here. The
second and third lines contain further known two-loop contributions
\cite{HSW0304}. The terms $\amu^{(\tilde{p}VS)}$ denote contributions from
diagrams where a vector boson $V$ and scalar $S$ couple to the muon line and
which involve a closed $\tilde{p}$-loop; $\amuSUTLferm$ and $\amuSUTLbos$
denote the difference of diagrams without SUSY particles between the MSSM and
the SM, arising from the different Higgs sectors. The dots denote known but
negligible terms computed in \cite{HSW0304}, the contributions computed
partially in \cite{FengLM06}, and the remaining, unknown contributions. For
analytical results
see the original references and \cite{review}.

\begin{figure}
\includegraphics[width=\columnwidth]{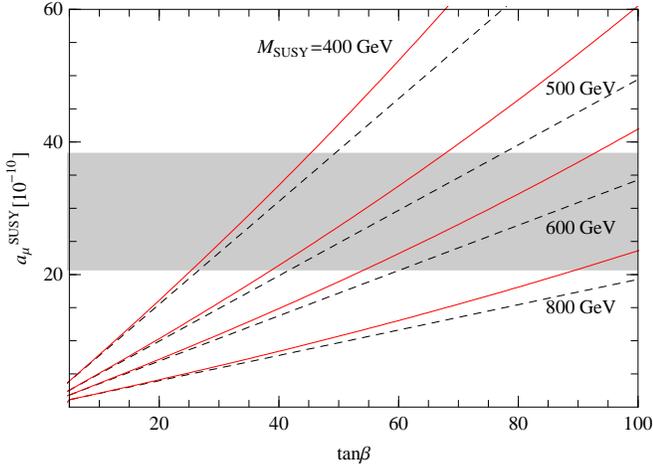}
\caption{$\amu$ as a function of $\tb$ for four different values 
of degenerate SUSY masses.  
Solid (red) lines: correct $\amu$ as in Eq.~(\ref{amuSUSYknown}).
Dashed (black) lines: $\amu$ without $\amu^{\rm SUSY,\Delta_\mu}$.  
Gray band: $1\sigma$ range of Eq.~(\ref{deviation}).
\label{fig}}
\end{figure}

In order to discuss the phenomenological impact of {our new}
contributions, we start by noting that
\begin{align}
\Delta_\mu &= -  0.0018 \tan\beta\;\mbox{sign}\, \mu 
\label{DeltamuMSUSY}
\end{align}
in the case where all SUSY masses are equal {and much larger than
  $M_W$.  Hence in the interesting region with $\tan\beta\sim 50$ the
  value of $\tan\beta$ extracted from $\amu^{\rm exp}$ will be off by
  roughly $10$\%, if $\amu^{\rm SUSY,\Delta_\mu}$ is omitted in
  Eq.~(\ref{ResultamuDeltamu}).  Fig.~\ref{fig} shows the impact of the
  new contribution on the dependence of $\amu$ on $\tb$.}

{Importantly, as a dimensionless quantity $\Delta_\mu$ does not
  decouple for arbitrarily large SUSY masses. For slight mass
  splittings, $\Delta_\mu$ can be even larger than in
  Eq.\ (\ref{DeltamuMSUSY}). 
For example, for $\tan\beta=50$ and $m_{L,\tilde{\mu}}=300$,
$m_{R,\tilde{\mu}}=500$, $M_2=650$, $\mu=800$~GeV and $M_1=M_2/2$ one
obtains a correction of $+14\%$ for $\amu^{\rm SUSY}$.}

{Among the SPS SUSY benchmark parameter points
  \cite{SPSDef} large effects are obtained at SPS~4 with
  $\tan\beta=50$ ($+8\%$), and at SPS~1b with $\tan\beta=30$ ($+6\%$).
In particular, for SPS~4, which is already experimentally disfavoured
by $\amu$,  the contribution rises from $\amu^{\rm SUSY}(\mbox{SPS 4})
= 49 \times 10^{-10}$ to $53\times10^{-10}$ by including the new
$\tan\beta$-enhanced correction. This corresponds to a rise of the
deviation from the experimental value from $2.2\sigma$ to
$2.6\sigma$. }

{In conclusion we have identified a new $\tb$-enhanced contribution
  to $\amu$ which first enters at the two-loop level. In scenarios with
  large values of $\tb$ the new term $\amu^{\rm SUSY,\Delta_\mu}$ alters
  the MSSM phenomenology by typically 10\%, but can have an even larger
  impact in certain regions of the parameter space. Our contribution is
  typically larger than the previously known supersymmetric two-loop
  corrections and should be included in global fits of electroweak
  precision data to the MSSM. 

This work is supported by the DFG--SFB/TR9 \emph{Computergest\"utzte
  Theoretische Teilchenphysik}, by BMBF grant 05 HT6VKB and by the EU Contract
No.~MRTN-CT-2006-035482, \lq\lq FLAVIAnet''.}

\end{document}